\documentclass[a4paper,10pt]{article}
\usepackage{amsmath,amssymb,epsfig,graphicx} 

\begin{document}

\title{GR@PPA Event Generator\footnote{
Talk presented at the 3rd Computational Particle Physics Workshop (CPP2010),
September 23-25, 2010, KEK, Tsukuba, Japan; 
submitted to Proceedings of Science.}}

\author{Shigeru Odaka\\
 High Energy Accelerator Research Organization (KEK)\\
 1-1 Oho, Tsukuba, Ibaraki 305-0801, Japan\\
 E-mail: \texttt{shigeru.odaka@kek.jp}}

\date{December 27, 2010}

\maketitle

\begin{abstract}
The history and the present status of the GR@PPA event generator are briefly reviewed.
The development of GR@PPA started in 2000 
in order to provide a framework of NLO event generators for hadron collision interactions. 
After the first release of the package (GR@PPA\_4b) in 2002, 
which supported four bottom quark production processes, 
many multi-particle production processes have been added to the package: 
GR@PPA\_ALL in 2004 and GR@PPA 2.7 in 2006. 
Now we are going to release a new package (GR@PPA 2.8), 
which supports single and double weak-boson production processes with an initial-state jet matching. 
Though the matrix elements presently included are still at the tree level, 
this new release is an important step towards constructing consistent NLO event generators.
\end{abstract}

\section{GR@PPA}

GR@PPA (GRace At Proton-Proton/Antiproton)\footnote{{\tt http://atlas.kek.jp/physics/nlo-wg/grappa.html}.}
\cite{Tsuno:2002ce,Tsuno:2006cu}
is a Monte Carlo event generator package to be used for simulating interactions 
at proton-proton and proton-antiproton collider experiments, 
such as those at CERN LHC and FNAL Tevatron.
It is an extension of the GRACE system \cite{Ishikawa:1993qr} to hadron collision interactions. 

The cross section of the production of a final state $A$ from collisions of two hadrons, 
$h_{1}$ and $h_{2}$, at a squared center-of-mass (cm) energy of $s$ is usually evaluated as:
\begin{eqnarray}
	{d\sigma_{h_{1}h_{2} \rightarrow A+X}(s) \over  d\Phi_{A}}
	= \sum_{a,b,i} \int_{0}^{1} dx_{1} \int_{0}^{1} dx_{2} 
	f_{h_{1} \rightarrow a}(x_{1},\mu_{F}^{2})f_{h_{2} \rightarrow b}(x_{2},\mu_{F}^{2}) 
	\delta({\hat s} - x_{1}x_{2}s)
	{ d{\hat \sigma}_{ab \rightarrow A_{i}}({\hat s}) \over d{\hat \Phi_{A_{i}}}},
\label{eq:xsec}
\end{eqnarray}
where $f_{h_{k} \rightarrow p}(x_{k},\mu_{F}^{2})$ is the parton distribution function (PDF) 
representing the existence probability of the parton $p$ (a light quark or a gluon) 
inside the hadron $h_{k}$ with a momentum fraction of $x_{k}$ 
at a certain energy scale of $\mu_{F}$ (factorization scale). 
The factor $d{\hat \sigma}_{ab \rightarrow A_{i}}({\hat s})/d{\hat \Phi_{A_{i}}}$ represents 
the differential cross section of the hard interaction 
which produces the final state $A_{i}$ from the collision of two partons, $a$ and $b$, 
perturbatively calculated according to fixed-order matrix elements (ME) 
at a squared cm energy of ${\hat s}$.
The final state $A$ may consist of several sub-states $A_{i}$ at the parton level; 
for instance, the "jet" production includes a variety of light quark and gluon productions.
Such a generalization is necessary in hadron collisions 
because it is experimentally hard to separate the sub-states.

Monte Carlo (MC) event generation is an MC integration 
in which sampling points (events) are generated with a probability proportional to the differential cross section. 
GRACE is a powerful tool to derive the differential cross section of the hard interaction,  
$d{\hat \sigma}_{ab \rightarrow A_{i}}({\hat s})/d{\hat \Phi_{A_{i}}}$ in  Eq.~(\ref{eq:xsec}), 
and to generate events according to it.
GR@PPA provides a mechanism to add the effects of the initial-state variation in the flavor and the momentum 
according to PDF and to make a generalization of the final state as symbolically described in Eq.~(\ref{eq:xsec}).
The summation in Eq.~(\ref{eq:xsec}) must sometimes be taken over more than tens of sub-processes.
It is not a good choice to prepare matrix elements for all sub-processes.
GR@PPA  simulates such sub-processes by applying charge and parity inversions, 
and flavor, mass and coupling exchanges to a limited number of base processes.
For example, all sub-processes for the $Z$ + 1 jet production are derived from only two base processes, 
$u{\bar u} \rightarrow Z + g$ and $ug \rightarrow Z + u$.
 
It is frequently required by users of MC event generators to mix up event samples of different processes 
according to the predicted production cross sections.
This requires additional cumbersome works.
GR@PPA has a mechanism to generate events of different processes simultaneously 
from the beginning of the development.
Event mixing is not necessary if users specify multiple processes in the initialization.
This is a very useful feature when we apply our jet matching method in GR@PPA.
Another feature implemented from the beginning is an interface to general-purpose event generators.
GR@PPA is a parton-level event generator.
It is necessary to apply additional simulations of hadronization and decays in order to generate realistic events 
which can be passed to detector simulations of experiments. 
General-purpose event generators, such as PYTHIA \cite{Sjostrand:2006za} 
and HERWIG \cite{Corcella:2000bw,Corcella:2002jc}, provide us with such a function.
In the early stage of the development, GR@PPA supported a dedicated interface to 
PYTHIA version 6.1 \cite{Sjostrand:2000wi}, 
while recent versions support the LHA user-process interface \cite{Boos:2001cv} 
commonly supported by various event generators.

\section{History of the development}

In January 2000, people from the Japanese ATLAS group and the Minami-Tateya group formed 
the NLO Working Group\footnote{{\tt http://atlas.kek.jp/physics/nlo-wg/index.html}.}.
The purpose of this working group is to develop next-to-leading order (NLO) event generators 
for hadron collision interactions based on the GRACE system.
The development of GR@PPA started as an activity of this working group.
In October 2000, a concept to integrate the GRACE system with PYTHIA was presented at the ACAT2000 workshop 
held at FNAL \cite{Sato:2001ae}.
According to this concept, we started the development of a leading-order (LO) event generator for four bottom-quark 
($b{\bar b}b{\bar b}$) production processes.
The generator included all processes within the Standard Model, such as those mediated by $Z$ boson(s) 
and the Higgs boson as well as those mediated by pure QCD interactions.
The development of four-body event generators was not a trivial task at that time.
The generator was named GR@PPA\_4b in February 2001, 
and GR@PPA\_4b version 1.0 was released in April 2002 \cite{Tsuno:2002ce}.
GR@PPA\_4b 2.0, released in April 2003, supported the LHA user-process interface.

After establishing GR@PPA\_4b, we started adding various multi-body production processes to the same framework.
In February 2004, we released a new version, GR@PPA\_All 2.6, in which in addition to the four-$b$ production 
many multi-body production processes, $W$ + jets up to 3 jets, $Z$ + jets up to 2 jets, diboson ($W^{+}W^{-}$, 
$ZW$, $ZZ$) productions and the top-quark pair production, were supported.
It must be emphasized that in these processes the decays of heavy particles ($W$, $Z$ and top quark) were included 
in the matrix elements for the event generation.
Namely, the "$W$ + 3 jets" is a five-body production and the "top-pair" is a six-body production process.
As a result, the phase-space effects and the spin correlations in the decays were reproduced exactly 
at the tree level in the generated events.
The most recent version of GR@PPA (GR@PPA 2.7) \cite{Tsuno:2006cu} was released in February 2006. 
Further multi-body processes, $W$ + 4 jets, $Z$ + 3 and 4 jets, diboson + 1 and 2 jets, top-pair + 1 jet, and 
QCD jets from 2 up to 4 jets, were added in this version.
Note that the "top-pair + 1 jet" is a seven-body production process.
In this version of the release, the GR@PPA framework and process-dependent routines are provided 
as separate packages, 
because the program size has grown very large for multi-body production processes.
Users are allowed to install only those processes which they are interested in.
The installation scripts have a function to recognize which processes are installed by the user.

The event generators included in GR@PPA have been verified through repeated comparisons with 
other event generators, especially for weak boson + jets production processes,
and the $W$ + jets event generators were utilized in a data analysis at Tevatron \cite{Tsuno:2004zz}.
However, through the analyses of Tevatron data, we have noticed that there is a fundamental difficulty 
in event generators based on matrix elements (ME) such as GR@PPA: 
a double counting problem between ME and parton showers (PS) or PDF for QCD radiation.
Parton showers are applied to simulate QCD radiation effects analytically evaluated in PDFs in Eq.~(\ref{eq:xsec}) 
and to simulate radiations from produced partons (jets).
The problem arises when we use matrix elements for those processes including jet(s) in the final state.
The phase space to which we can naively apply ME-based event generators is much restricted 
if we want to safely avoid the double counting.
Namely, we have to require a sufficiently large separation between the jets.
We have to add an appropriate matching mechanism in order to construct event generators 
which provide us with reasonable simulations in the entire phase space.

\section{ME-PS matching and GR@PPA 2.8}

The purpose of the NLO working group is to develop NLO event generators. 
It is necessary to deal with those processes including at least one QCD radiation in NLO event generators.
The development of GR@PPA aims at providing an appropriate platform of event generator programs.
Therefore, the double counting is a problem that we have to solve in GR@PPA.

The double counting between ME and PS is now a well known problem.
Several solutions have been introduced and implemented in practical event generators, 
such as the ME correction in PYTHIA \cite{Miu:1998ju} and HERWIG \cite{Seymour:1994we,Seymour:1994df}, 
and the CKKW method \cite{Catani:2001cc} implemented in Sherpa \cite{Gleisberg:2008ta}. 
The MLM prescription in AlpGen \cite{Mangano:2002ea} can be considered as an alternative to the CKKW method.
These are the solutions for leading-order (LO) event generators.
For full NLO event generation, a subtraction method is applied in MC@NLO \cite{Frixione:2002ik,Frixione:2008ym} 
and a suppression method is used in POWHEG \cite{Nason:2004rx,Alioli:2010xd}.

We have proposed another solution, the limited leading-log (LLL) subtraction method 
\cite{Kurihara:2002ne,Odaka:2007gu}, to the problem.
The divergent leading-logarithmic (LL) terms are subtracted from the matrix elements 
of radiative (1-jet) processes in this method. 
This is similar to the subtraction method applied in MC@NLO.
However, the subtraction is applied in a limited phase space in our method, 
while there is no such limitation in the method of MC@NLO.
In any case, the "1-jet" processes become finite after the subtraction.
The subtracted terms are equal to the leading terms considered in PS applied to non-radiative (0-jet) processes. 
The duplication of these terms is the source of the double counting.
Since the duplicated terms are subtracted from "1-jet" MEs, 
we can achieve an appropriate matching by combining them with "0-jet" processes to which a PS is applied.

Since the LL subtraction is unphysical, 
we may have negative cross sections in some phase space after the subtraction, 
leading to the generation of some negative-weight events.
Generated events have a unit weight, $+1$ or $-1$, in GR@PPA and MC@NLO.
Physical distributions can be obtained by subtracting the number of negative-weight 
events from that of positive-weight events in each histogram bin, 
after adding "0-jet" and "1-jet" processes.
The generation of negative-weight events is not a serious problem 
unless their fraction becomes large.

The factorization of the perturbative QCD radiation effects (QCD evolution) stops at 
the factorization scale ($\mu_{F}$) in PDFs.
This scale is usually taken to be equal to a typical energy scale ($s$ or $|t|$) of the interested interaction, 
because non-leading contributions ignored in the factorization become sizable around this scale.
Parton shower (PS) is in principle an MC implementation of the QCD evolution considered in PDF.
Therefore, it is natural to limit the PS radiation also at $\mu_{F}$.
The LL subtraction is a rejection of the duplication between ME and PS.
If the evolution in PDF and the PS radiation are limited, the LL subtraction must also be limited by the same energy scale.
The subtraction is applied in the phase space limited by $\mu_{F}$ in the LLL subtraction.
The LL radiation effects harder than $\mu_{F}$ remain in "1-jet" processes.
Therefore, we can expect that $\mu_{F}$ dependences must be cancelled  with an accuracy at the LL order 
by adding "0-jet" and LLL-subtracted "1-jet" processes. 
We have shown that such a cancellation is actually realized \cite{Odaka:2007gu}.

Parton showers (PS) play a crucial role in the jet matching in event generators.
We have developed custom-made parton showers \cite{Kurihara:2002ne} 
in order to ensure the performance of our method.
They are based on the most naive definition of the Sudakov form factor at the leading order.
The initial-state PS, named QCDPS, is a forward evolution PS.
It strictly reproduces the QCD evolution in leading-order PDFs. 
QCDPS overcomes the efficiency problem that forward-evolution parton showers generally suffer from, 
by adopting an $x$-deterministic evolution technique. 

Parton showers simulate not only the longitudinal momentum branches but also transverse activities of radiations.
Their cumulative effects can be observed as physical quantities, such as the transverse recoil momentum 
of weak bosons.
We have shown that the definition of PS branch kinematics is important to achieve a good matching 
in the transverse recoil, 
and introduced a new definition named "$p_{T}$-prefixed branch kinematics" \cite{Odaka:2007gu}.
The simulation adopting this definition almost perfectly reproduces the transverse momentum spectrum 
of $Z$ bosons measured at Tevatron \cite{Odaka:2009qf}. 
In addition to QCDPS, we have developed a backward-evolution initial-state PS 
with the same branch kinematics (QCDPSb).
A final-state PS (QCDPSf) has also been developed for completeness, but is still in an experimental phase.

A new version of GR@PPA (GR@PPA 2.8) is now almost ready to release, 
in which the ME-PS matching method described above is applied to weak-boson production processes.
Supported are the single $W$ and $Z$ production processes and the diboson ($W^{+}W^{-}$, 
$ZW$ and $ZZ$) production processes.
Used matrix elements are still at the leading order.

\begin{figure}
\begin{center}
\includegraphics[scale=0.9]{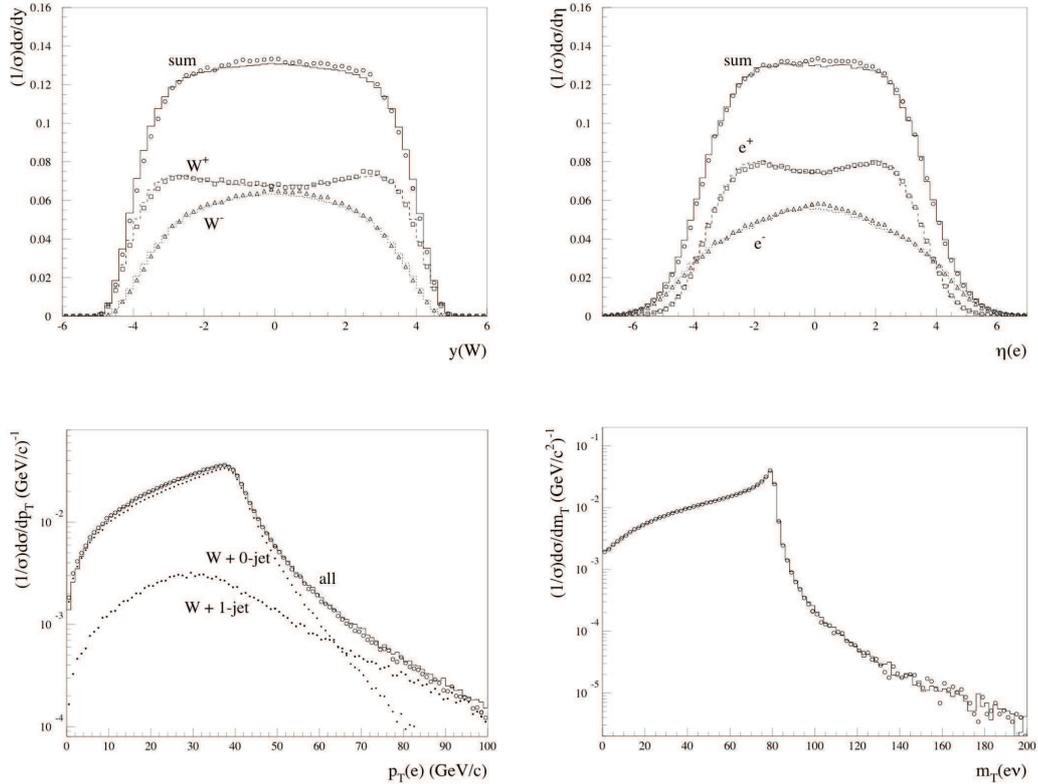}
\caption{\label{fig:w}
Some properties of $W$ bosons and decay electrons in the single $W$-boson production process 
at the LHC condition, $pp$ collisions at a cm energy of 14 TeV:
the rapidity of $W$ (top-left), the pseudorapidity of electrons (top-right), 
the $p_{T}$ of electrons (bottom-left), and the transverse mass calculated from $p_{T}$ of 
electrons and neutrinos (bottom-right).
The predictions from GR@PPA 2.8 (plots) are compared with those from PYTHIA with the "new" PS 
(histograms).
}
\end{center}
\end{figure}

\begin{figure}
\begin{center}
\includegraphics[scale=0.65]{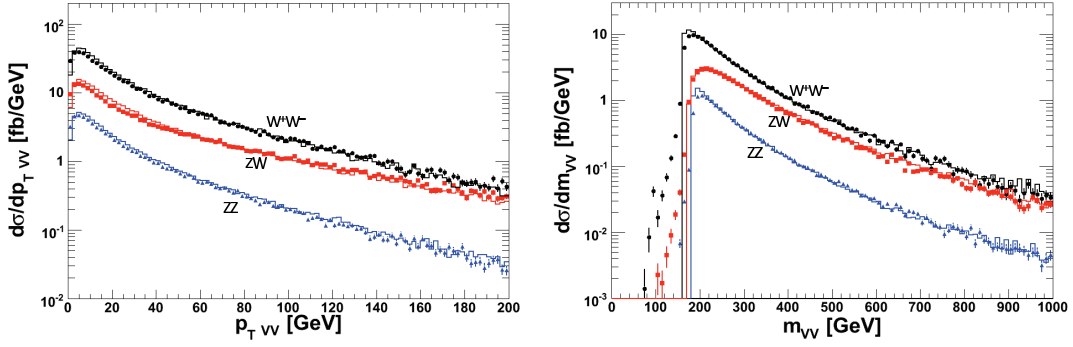}
\caption{\label{fig:vv}
The $p_{T}$ and invariant mass spectra of the diboson system in the diboson production processes 
at the LHC condition, $pp$ collisions at a cm energy of 14 TeV.
GR@PPA simulations (plots) are compared with predictions from MC@NLO (histograms).
}
\end{center}
\end{figure}

As already mentioned, the generated events almost perfectly reproduces the $p_{T}$ spectrum 
of $Z$ bosons at Tevatron. 
Here we show some results on the $W$-boson production at the LHC condition,  
$pp$ collisions at a cm energy of 14 TeV, in Fig.~\ref{fig:w}.
Some properties of $W$ bosons and decay electrons are compared with those simulated 
by PYTHIA \cite{Sjostrand:2006za}.
We expect that the PYTHIA simulation is at the same level as the present GR@PPA simulation 
concerning the $W$-boson production, except for the overall normalization.
Actually we can see a good agreement between the two simulations in Fig.~\ref{fig:w},
where relative shapes of the distributions are compared.
As for the diboson production, the predictions for the LHC condition are in reasonable agreement 
with those from MC@NLO \cite{Frixione:2008ym} as shown in Fig.~\ref{fig:vv}.
The disagreements can be attributed to the absence of virtual corrections in GR@PPA 
and the zero decay widths of weak bosons assumed in MC@NLO.
Weak bosons are produced according to the on-shell approximation in these simulations of MC@NLO.
Finite decay widths and the spin correlations can be simulated in MC@NLO only for the $W^{+}W^{-}$ 
production process,
while these properties are simulated in GR@PPA for all processes.

\section{Summary and prospects}

\begin{figure}
\begin{center}
\includegraphics[scale=0.65]{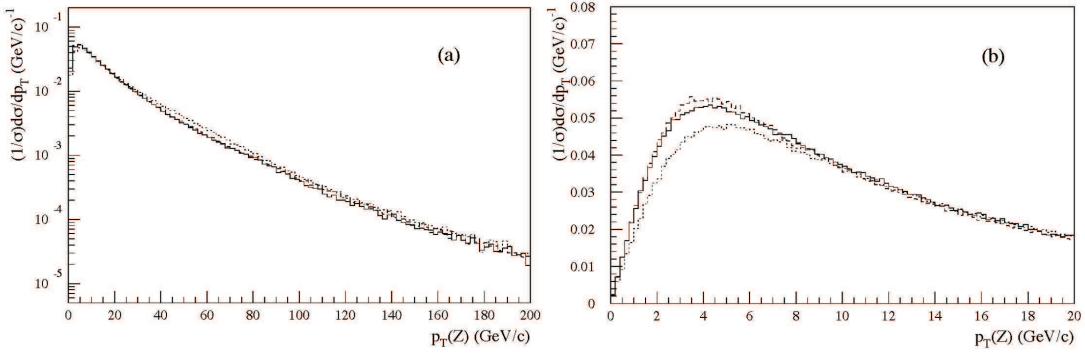}
\caption{\label{fig:zpt-lhc}
The $p_{T}$ spectrum of $Z$ bosons in the single $Z$ production process at LHC, 
$pp$ collisions at 14 TeV.
The prediction from GR@PPA 2.8 (solid) is compared with those from PYTHIA with the new PS 
(dashed) and HERWIG (dotted). 
The distributions up to 200 GeV/$c$ are shown in (a) in order to compare the high-$p_{T}$ spectra, 
while the low-$p_{T}$ spectra up to 20 GeV/$c$ are compared in (b).
}
\end{center}
\end{figure}

The NLO working group was formed in January 2000, aiming at developing NLO 
event generators for hadron-collision interactions based on the GRACE system.
GR@PPA has been developed as an activity of this working group, 
in order to establish a framework of the event generator program. 
Though the implemented matrix elements are still at the tree level, 
many multi-body (multi-jet) production processes are supported in the GR@PPA packages
released so far. 

The ME-PS matching is a feature that practical NLO event generators have to support.
We have introduced the LLL subtraction method for initial-state jet matching 
to smoothly combine "0-jet" and "1-jet" processes, and applied it to weak-boson production processes.
The generated events surprisingly well reproduce the $p_{T}$ spectrum of $Z$ bosons measured at Tevatron.
A new version of the GR@PPA package (GR@PPA 2.8) is now almost 
ready to release\footnote{GR@PPA 2.8 was released on Nov. 17, 2010. See the GR@PPA Web page.}.
This version supports the single $W$ and $Z$ production processes and diboson ($W^{+}W^{-}$, 
$ZW$ and $ZZ$) production processes.
The LLL subtraction can be applied to combine them with "1-jet" associate processes, 
to reproduce the recoil of the weak-boson system due to QCD radiations in the entire phase space.
Leading-order parton shower programs that we have developed are also included 
in order to ensure the performance of our matching method.
This release is an important step towards our final goal.

As the next step of the development, 
we are going to apply the LLL subtraction method to the di-photon production process. 
This is an important background process for light Higgs-boson searches, 
and the contribution of jet-associate processes is known to be very large.
Since some of the jet-associate sub-processes have QED divergences in the final state, 
we need to extend the QCD LLL subtraction for the initial state, that we have developed so far, 
to QED and the final state. 
Another subject that we are going to attack is to construct full NLO event generators.
Since we can evaluate loop integrations efficiently in the framework of GRACE, 
it is straightforward to convert present event generators in GR@PPA 2.8 to NLO ones \cite{Fujimoto:2008zz}, 
once an appropriate theoretical formalism is established. 

We also have a plan of developments in technical aspects.
The present GR@PPA coding still depend on human manipulations in many parts.
The coding becomes more and more complicated as the particle multiplicity becomes larger, 
and the chance to make mistakes becomes more frequent.
It is desired to pursue further automatization in order to reduce such mistakes.
Some of the developments are underway in this course.

Continuous physics runs started at LHC early in this year. 
The accumulated luminosity is increasing exponentially according to the rapid improvements 
of the accelerator performance.
Figure~\ref{fig:zpt-lhc} shows the $p_{T}$ spectrum of $Z$ bosons expected 
for the design condition of LHC, $pp$ collisions at 14 TeV.
Predictions from GR@PPA, PYTHIA with the new PS and HERWIG are compared in the figure.
Though unfortunately the predictions are quite similar to each other, 
there still exist some substantial differences between them. 
It is easy to produce a similar figure for the present LHC condition, $pp$ collisions at 7 TeV.
We expect that we can plot measurement results on it, soon.

\end{document}